\documentclass[aps,twocolumn,secnumarabic,balancelastpage,amsmath,amssymb,nofootinbib,floatfix]{revtex4-2}

\usepackage{graphicx}      
\usepackage[font = footnotesize]{caption}
\usepackage{tabularx}
\usepackage{siunitx}
\usepackage{bm}
\usepackage{amsmath}
\usepackage[colorlinks=true]{hyperref}  

\newcolumntype{P}[1]{>{\centering\arraybackslash}p{#1}}

\DeclareMathOperator{\tr}{tr}

\begin{document}
	
\title{Zeptonewton and Attotesla per Centimeter Metrology With Coupled Oscillators}
\author{Ian Bouche}
\email{ianbo@bu.edu}
\author{Josh Javor}
\author{Abhishek Som}
\author{David K. Campbell}
\author{David J. Bishop}
\date{\today}
\affiliation{Bishop Lab, Department of Physics, Boston University \\
Boston, MA 02215, United States}

\begin{abstract}
	We present the coupled oscillator: a new mechanism for signal amplification with widespread application in metrology. We introduce the mechanical theory of this framework, and support it by way of simulations. We present a particular implementation of coupled oscillators: a microelectromechanical system (MEMS) that uses one large ($\sim$\SI{100}{\milli\meter}) N52 magnet coupled magnetically to a small ($\sim$\SI{0.25}{\milli\meter}), oscillating N52 magnet, providing a force resolution of \SI{200}{\zepto\newton} measured over \SI{1}{\second} in a noiseless environment. We show that the same system is able to resolve magnetic gradients of \SI{130}{\atto\tesla/\centi\meter} at a single point (within \SI{500}{\micro\meter}). This technology therefore has the potential to revolutionize force and magnetic gradient sensing, including high-impact areas such cardiac and brain imaging.
\end{abstract}

\maketitle

\textbf{We introduce and simulate a new technology for detecting small forces and magnetic gradients, leveraging nonlinear effects as an oscillator approaches a singularity in its potential. This \textit{coupled oscillator} framework achieves sensitivities that are highly competitive with the state of the art (e.g. atomic force microscopy for forces, superconducting quantum interference devices for magnetic gradients) while retaining ease of manufacturing and an accessible price point. The example simulated in this paper achieves 200$\,$zN and 130$\,$aT/cm resolutions using 0.25$\,$mm cubical magnet as an oscillator, with signals averaged over 1$\,$s.}

\section{Introduction}
	Precision metrology is one of the most widespread applications of microelectromechanical systems (MEMS) because powerful nanofabrication techniques and beneficial scaling laws create high sensitivity to small inputs. In recent years, new MEMS technologies have emerged that use a distance-dependent force between an oscillator and another object for precise metrology (\cite{Josh's analysis paper}, \cite{Zeptometer metrology paper}, \cite{Matthias' paper}). 
	
	This technology, deemed here as the coupled oscillator, has shown extraordinary promise in force and magnetic gradient metrology. Rather than measuring deflection of a component under a force (as in some traditional cantilever-based force metrology frameworks), a coupled oscillator transduces the input force to a shift in oscillation frequency. This is key for high-sensitivity measurement: by working in frequency space, we can take advantage of an enhanced resolution to small changes thanks to high-precision frequency counters and atomic clock standards. Further, transducing the signal to an oscillation allows for the use of powerful noise-reduction schemes such as phase-locking amplification.
	
	In what follows, we will describe the most general form of a coupled oscillator and show how we can use it to measure small forces. Then, we will propose a special case of this framework that uses magnetic attraction as the coupling force and show using simulations that it can theoretically provide sensitivities to \SI{200}{\zepto\newton} and \SI{460}{\yocto\newton} forces with \SI{1}{\second} and \SI{100}{\second} time gates, respectively. These numbers are achieved using commercially available frequency counters and rubidium clocks, but recent developments in atomic clock technology could grant relative frequency measurement uncertainties of \SI{e-18}{} \cite{Precision frequency measurement with ultracold atoms}, which would theoretically grant 6 to 8 orders of magnitude of additional sensitivity over the reported numbers at similar gate times.
	
	We will see that this framework can achieve higher sensitivity with oscillators that have lower masses and softer springs. Therefore, this technology will lend itself to MEMS platforms due to favorable scaling laws \cite{MEMS textbook}. In this way, coupled oscillator technology naturally integrates with nanometer-range interactions such as the Casimir effect (\cite{Josh's analysis paper}, \cite{Zeptometer metrology paper}, \cite{Matthias' paper}), and offers a method to directly and precisely measure these quantum interactions.
	
	We will also present this system as a platform for the measurement of small magnetic sources that is highly competitive with leading technologies such as atomic magnetometers and superconducting quantum interference devices. There is a strong demand for increased magnetic precision in experimental physics disciplines that such as cold dark matter searches \cite{ADMX} and transition-edge sensor systems \cite{Transition-edge sensors} including various observatories and astronomy experiments (e.g. see source \cite{SCUBA-2 has TES}). Further, one of the premier areas of interest for precision magnetometry and gradiometry is biomedical imaging. The motion of charged ions through the heart during pumping produces magnetic fields near the chest of $\sim\negthickspace\SI{100}{\pico\tesla}$. This signal, however, is drowned in noise from the Earth's magnetic field, whose fluctuations are of order $\sim\negthickspace\SI{200}{\pico\tesla}$ \cite{Geomagnetic noise paper}. For this reason, leading technologies sensitive enough to image the heart's magnetic fields such as atomic magnetometers or superconducting quantum interference devices generally require rooms dedicated to magnetic shielding, which significantly detriments the accessibility and price point of those techniques. However, the gradient of Earth's magnetic field and its noise are both significantly below the gradients present near the human heart \cite{Feasibility of magnetocardiography with gradients paper}. In this way, technology that is sensitive only to gradients of the magnetic field (such as the design proposed in this work) may offer an inexpensive, accessible and uncomplicated cardiac imaging method. The coupled oscillator system described in this paper will be shown to be sensitive to magnetic gradients of \SI{130}{\atto\tesla/\cm} and \SI{310}{\zepto\tesla/\cm} at a single point (within \SI{500}{\micro\meter}) for respective gate times of \SI{1}{\second} and \SI{100}{\second} under noiseless conditions. These resolutions are well under the threshold required for biomedical sensing of the heart. Coupled oscillator gradiometry may also be applied to other biomedical sources such as sensing of the brain's magnetic signals, which are 2-3 orders of magnitude weaker than those of the heart \cite{Encephalography explainer} and whose measurement similarly relies on shielding with presently available technologies.
	
\section{Coupled Oscillator Framework}
\begin{figure}[!h]
		\includegraphics[width=\linewidth]{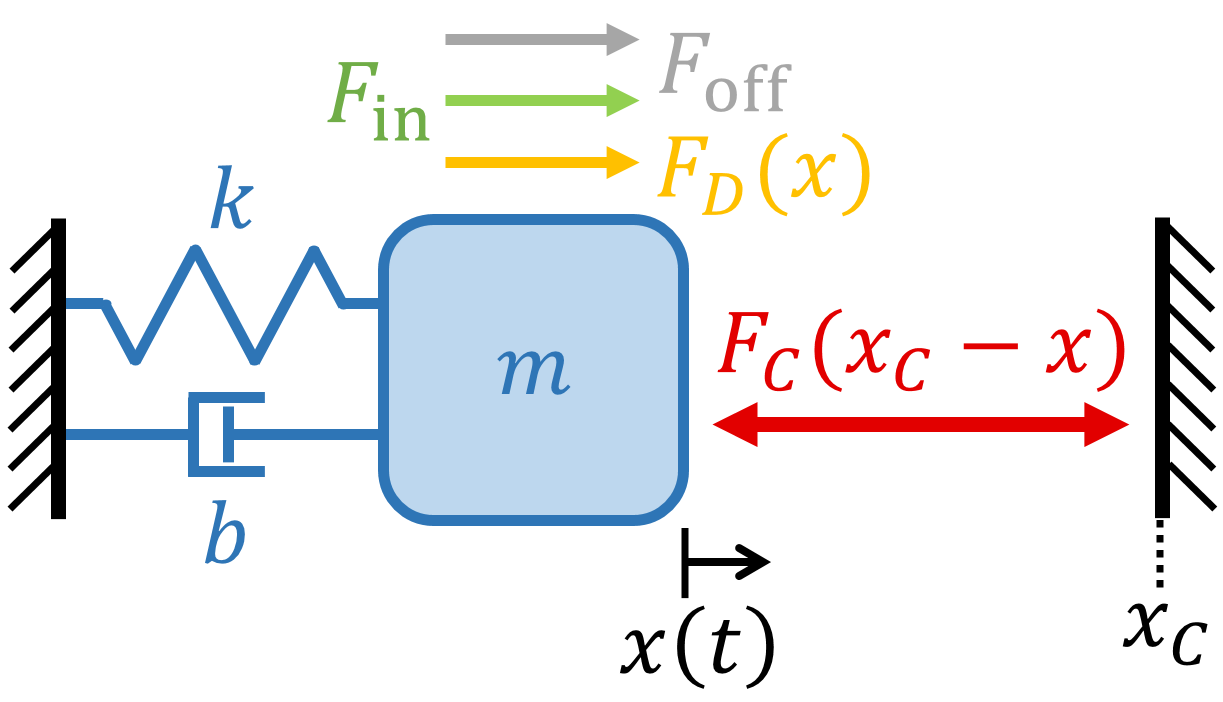}
		\caption{Coupled oscillator framework. A damped harmonic oscillator (blue) is driven by a phase-locked sinusoidal force $F_D(x)$. This force is linked to the position of the oscillator through phase locking. The oscillator also experiences a coupling force $F_C$, a constant input force $F_{\text{in}}$ which we want to measure, and a constant offset force $F_\text{off}$.}
		\label{fig:Coupled Oscillator Diagram}
	\end{figure}

	Consider a damped harmonic oscillator with the spring equilibrium position at $x = 0$ and various added forces (fig. \ref{fig:Coupled Oscillator Diagram}):
	\begin{itemize}
		\item \textbf{Coupling force} $F_C$: Force felt by the oscillator due to its proximity to a special point called the coupling wall, located at $x = x_C$. Examples of coupling forces could be magnetic forces (if, for example, the oscillator and coupling wall are magnets) or the Casimir force (\cite{Josh's analysis paper}, \cite{Zeptometer metrology paper}, \cite{Matthias' paper}). Positive values of $F_C$ are taken to represent an attractive force while negative values represent repulsion. In this work, this force is always written as a function of the distance between the oscillator and the coupling wall, and never as a function of absolute position in our coordinate system.
		
		\item \textbf{Driving force} $F_D$: Time-dependent sinusoidal driving force with constant amplitude. This force is phase-locked to lead the motion of the oscillator by \SI{90}{\degree} such that, as the oscillator's resonant frequency changes during measurement, the driving frequency will adjust to match it. More precisely, the system will always approach the undamped resonant frequency of the system $\omega_0 = \sqrt{\frac{k}{m}}$ rather than the true resonant frequency, which depends on damping. This critical feedback loop is modeled and demonstrated in the appendix. Further, this behavior of phase-locked loops has been previously shown in references \cite{Phase-locked-loops to maintain resonance 1}, \cite{Phase-locked-loops to maintain resonance 2}.
		
		\item \textbf{Input force} $F_\text{in}$: The force that we want to measure. This force will change the equilibrium position of the oscillator.
		
		\item \textbf{Offset force} $F_\text{off}$: Constant force to cancel the coupling force at $x = 0$. Therefore, $F_\text{off} = -F_C(x_C)$ always. This force ensures that the system's equilibrium is at $x = 0$ when there is no input force, regardless of the baseline distance to the coupling wall $x_C$. This allows us to pick the value of $x_C$ freely to optimize performance.
	\end{itemize}
	
	\subsection*{Measurement mechanism}
	In the following derivation, we will denote locations in space in two ways:
	\begin{itemize}
		\item Position $x$, where $x = 0$ represents the equilibrium position of the spring, and $x$ increases as we approach the coupling wall located at $x = x_C$.
		\item Coupling wall distance $d$, where $d = 0$ represents the position of the coupling wall ($x_C$) and distance increases as we move from the coupling wall towards the equilibrium position of the spring.
	\end{itemize}
	We can therefore convert between any position $x_\text{label}$ and distance $d_\text{label}$ with a simple transformation.
	\begin{equation*}
		d_\text{label} = x_C - x_\text{label}
	\end{equation*}
	The equation of motion of this system is then
	\begin{equation*}
		m\ddot{x} + b\dot{x} + kx = F_D(t) + F_C(x_C - x) + F_\text{in} + F_\text{off}
	\end{equation*}
	We write the driving force as being time-dependent for clarity, but it is fully determined by the position of the oscillator through phase-locking rather than the time.
	This system has a stable equilibrium position $x_\text{eq}$ and distance $d_\text{eq}$, defined implicitly by setting $\dot{x}$, $\ddot{x} = 0$ and neglecting the driving force
	\begin{equation}
		kx_\text{eq} = F_C(d_\text{eq}) + F_\text{in} + F_\text{off}
		\label{eqn:Equilibrium position definition}
	\end{equation}
	We introduce the displacement from equilibrium $\Delta x$:
	\begin{equation*}
		\Delta x \equiv x - x_\text{eq}
	\end{equation*}
	And we rewrite the equation of motion
	\begin{align}
		m\ddot{\Delta x} + b\dot{\Delta x} + & k\Delta x + kx_\text{eq} = \notag \\ 
		& F_D(t) + F_C(d_\text{eq} - \Delta x) + F_\text{in} + F_\text{off} \notag
	\end{align}
	We Taylor-expand the coupling force about the equilibrium distance
	\begin{align}
		& m\ddot{\Delta x} + b\dot{\Delta x} + k\Delta x + kx_\text{eq} = \notag \\
		& F_D(t) + F_C(d_\text{eq}) - \Delta xF_C'(d_\text{eq}) + \mathcal{O}(\Delta x^2) + F_\text{in} + F_\text{off} \notag
	\end{align}
	By rearranging terms and using eq. \ref{eqn:Equilibrium position definition}, we obtain
	\begin{equation*}
		m\ddot{\Delta x} + b\dot{\Delta x} + (k + F_C'(d_\text{eq}))\Delta x + \mathcal{O}(\Delta x^2) = F_D(t)
	\end{equation*}
	In order to ignore the $\mathcal{O}(\Delta x^2)$ anharmonic terms, the linear order term $(k + F_C'(d_\text{eq}))\Delta x$ must exceed them significantly. By comparing the $n$th term of the series with the linear term, this condition can be written as
	\begin{equation}
		\Delta x \ll \sqrt[n - 1]{n!\frac{k + F_C'(d_\text{eq})}{F_C^{(n)}(d_\text{eq})}} \ \ \ \ 
		\forall n = 2, 3, 4, ...
		\label{eqn:Linearity condition}
	\end{equation}
	We can see that the expression shown here is impossible to satisfy if $F_C'(d_\text{eq}) = -k$ which may occur for attractive coupling forces ($F_C > 0$). We call the value of $d_\text{eq}$ at which this happens the stiction distance $d_\text{stic}$, defined implicitly by
	\begin{equation*}
		F_C'(d_\text{stic}) \equiv -k
	\end{equation*}
	If we study the limit $d_\text{eq} \rightarrow d_\text{stic}$, we can then rewrite expression \ref{eqn:Linearity condition} by Taylor-expanding about $d_\text{eq} = d_\text{stic}$ and remove terms of order $\mathcal{O}((d_\text{eq} - d_\text{stic})^2)$ as
	\begin{equation*}
		\Delta x \ll \sqrt[n - 1]{n!\frac{k + F_C'(d_\text{stic}) + (d_\text{eq} - d_\text{stic})F_C''(d_\text{stic})}{F_C^{(n)}(d_\text{stic}) + (d_\text{eq} - d_\text{stic})F_C^{(n + 1)}(d_\text{stic})}}
	\end{equation*}
	Since $F_C'(d_\text{stic}) \equiv -k$, the numerator simplifies. Expanding our fraction in powers of $d_\text{eq} - d_\text{stic}$ and again neglecting terms of order $\mathcal{O}((d_\text{eq} - d_\text{stic})^2)$, our expression becomes
	\begin{equation*}
		\Delta x \ll \sqrt[n - 1]{n!(d_\text{eq} - d_\text{stic})\frac{F_C''(d_\text{stic})}{F_C^{(n)}(d_\text{stic})}}
	\end{equation*}
	In our limit $d_\text{eq} \rightarrow d_\text{stic}$, the $n = 2$ expression will always be the smallest (due to the linear dependence on $d_\text{eq} - d_\text{stic}$ at $n = 2$ vs root dependence for $n \neq 2$), so it will form the most stringent upper bound on $\Delta x$. Therefore we can set $n = 2$ without loss of generality, and neglect the factor of $2$ to obtain a simplified linearity condition
	\begin{equation}
		\Delta x \ll d_\text{eq} - d_\text{stic}
		\label{eqn:Simple linearity condition}
	\end{equation}
	So, by ensuring small oscillations to meet condition \ref{eqn:Simple linearity condition}, we can ignore the higher-order terms in the equation of motion, leading to a driven, damped harmonic oscillator with a modified spring constant $k_\text{shifted} = k + F_C'(d_\text{eq})$
	\begin{equation*}
		m\ddot{\Delta x} + b\dot{\Delta x} + k_\text{shifted}\Delta x = F_D(t)
	\end{equation*}
	The resonant frequency of this system is known to be
	\begin{align}
		f_\text{shifted} &= \frac{1}{2\pi}\sqrt{\frac{k_\text{shifted}}{m} - \frac{1}{2}\left(\frac{b}{m}\right)^2} \notag \\
		    &= \frac{1}{2\pi}\sqrt{\frac{k + F_C'(d_\text{eq})}{m} - \frac{1}{2}\left(\frac{b}{m}\right)^2} \notag \\
		    &= \frac{1}{2\pi}\sqrt{\frac{(1 - 1/2Q^2)k + F_C'(d_\text{eq})}{m}} \notag
	\end{align}
	Where we rewrote $b$ in terms of the quality factor $Q = \sqrt{km}/b$. In the high-quality limit ($Q\rightarrow\infty$), the shifted frequency is
	\begin{equation}
		f_\text{shifted} = \sqrt{1 + \frac{F_C'(d_\text{eq})}{k}}f_0
		\label{eqn:Shifted Frequency}
	\end{equation}
	where $f_0 = \frac{1}{2\pi}\sqrt{\frac{k}{m}}$ is the undamped harmonic oscillator resonant frequency.
	
	We should note that even for larger damping and lower quality factors, our phase-locking feedback loop will not approach the true resonant frequency, but rather the undamped resonant frequency, always. This is shown and explored in the appendix.
	
	Equation \ref{eqn:Shifted Frequency} then shows how the coupling force manifests as a change in the system: if the oscillator equilibrium is moved towards the coupling wall, its resonance frequency will either decrease (if the coupling is attractive) or increase (if it is repulsive). The agent that causes this approach is $F_\text{in}$, meaning the force we want to measure induces a shift in the system's resonant frequency. This is the mechanism by which forces we wish to measure are turned into frequency shift signals. It is important to note that $d_\text{eq}$ is related to $F_\text{in}$ in a nonlinear way that depends on the form of $F_C(d)$. The exact relationship can be found by solving equation \ref{eqn:Equilibrium position definition}.
	
\begin{figure}[!h]
		\includegraphics[width=\linewidth]{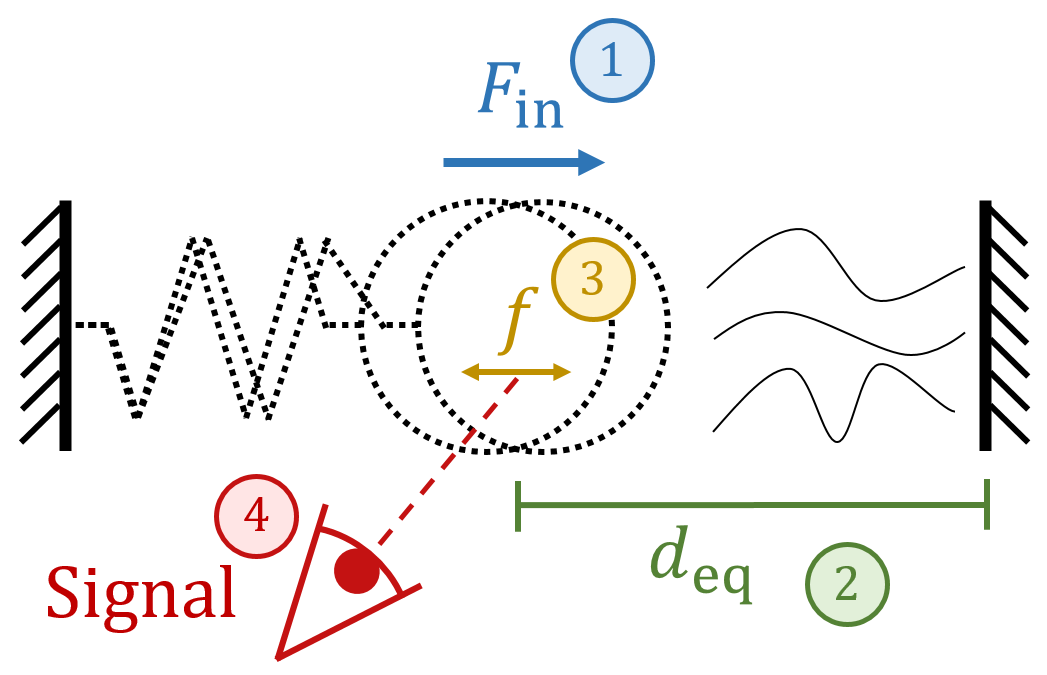}
		\caption{Overview of the four steps of the force detection mechanism of a coupled oscillator.}
		\label{fig:Signal Pathway}
	\end{figure}

	We can express the measurement process as a sequence of steps, visualized in figure \ref{fig:Signal Pathway}:
	\begin{enumerate}
		\item The input force acts on the oscillator.
		
		\item This creates a shift in the equilibrium position of the oscillator, moving it closer to or further from the coupled wall.
		
		\item The new distance to the coupling wall changes the effective spring constant which in turn creates a change in the resonant frequency as per equation \ref{eqn:Shifted Frequency}. Due to phase locking of the driving force, this eventually leads the driving frequency to match the undamped resonant frequency.
		
		\item This change in frequency is measured using frequency counters, and this is read as the signal.
	\end{enumerate}
	
	\subsection*{Stiction}
\begin{figure}[!h]
		\includegraphics[width=\linewidth]{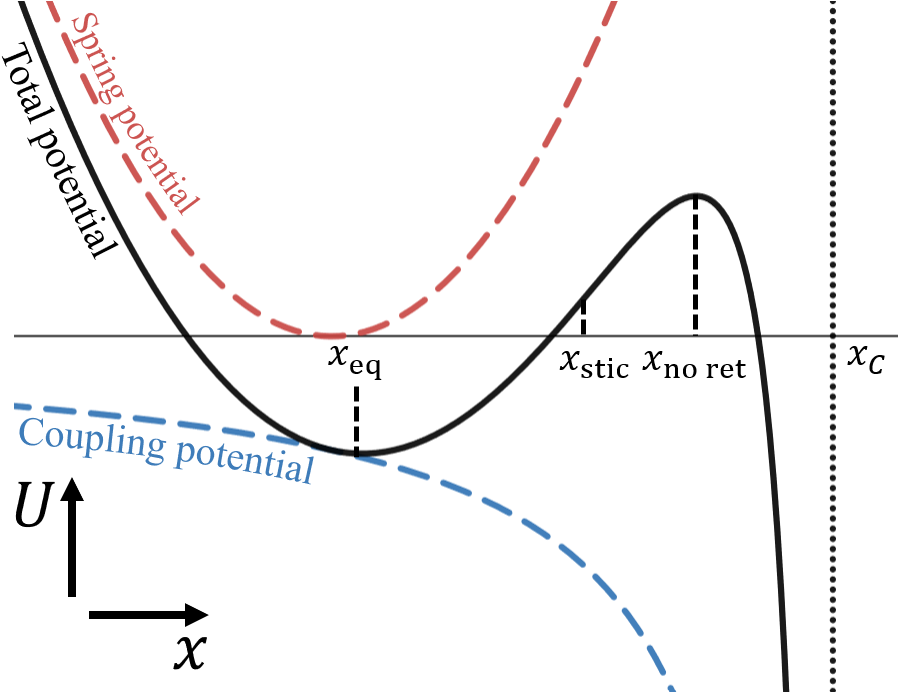}
		\caption{Quasistatic potential experienced by the oscillator with an attractive coupling force. The potentials $U(x)$ for the spring and coupling force add to create the landscape shown. If the oscillator passes the point of no return $x_\text{no ret}$, it will necessarily collide with the coupling wall. If the spring potential were moved towards the coupling wall, the curvature at equilibrium would decrease until it is zero when $x_\text{eq} = x_\text{stic}$.}
		\label{fig:Quasistatic Potential}
	\end{figure}

	If at any point $F_C'(d) \leq -k$, the system will experience a breakdown in oscillation and a runaway attraction towards the coupling wall. This event is called stiction, and it can only occur when condition \ref{eqn:Simple linearity condition} is violated. Stiction deems the system unusable and can be catastrophically damaging for some fragile implementations. The stiction position $x_\text{stic} = x_C - d_\text{stic}$ can be seen in figure \ref{fig:Quasistatic Potential} in the potential experienced by the oscillator. 
	
	It is important to distinguish the stiction position $x_\text{stic}$ from the point of no return of the oscillator $x_\text{no ret}$. $x_\text{stic}$ is the highest value of $x_\text{eq}$ that avoids stiction, while $x_\text{no ret}$ is the highest value of $x$ that avoids stiction. Although violating the linearity condition \ref{eqn:Simple linearity condition}, it is possible to have $x > x_\text{stic}$ without stiction occurring, but not $x > x_\text{no ret}$. Further, as we move the equilibrium position towards the coupling wall (for instance, by considering stronger input forces), $x_\text{stic}$ will not move, but $x_\text{no ret}$ will. In figure \ref{fig:Quasistatic Potential}, $x_\text{stic}$ corresponds to the inflection point of the potential energy while $x_\text{no ret}$ corresponds to a local maximum.
	
	While stiction is generally dangerous, operating near it can also grant extraordinary sensitivity, as we will see in the analysis of the gain.
	
	\subsection*{Gain of coupled oscillators}
	Next we will study the amplification granted by the coupled oscillator framework. Consider the measurement gain $G$:
	\begin{equation*}
		G \equiv \frac{1}{\Delta F_\text{in}}
	\end{equation*}
	Where $\Delta F_\text{in}$ represents the smallest change in input force we can measure. By considering the steps of the measurement, we can write the gain as:
	\begin{equation*}
		G = \frac{1}{\Delta f_\text{shifted}} \frac{\Delta f_\text{shifted}}{\Delta d_\text{eq}} \frac{\Delta d_\text{eq}}{\Delta F_\text{in}}
	\end{equation*}
	
	The first factor is the system's ability to to detect frequency shifts. This is determined by the precision of the frequency counters and the time gate settings used to detect frequency signals. We will write this as $\frac{1}{\Delta f_\text{min}}$, where the frequency resolution $\Delta f_\text{min}$ represents the smallest frequency shift we can detect.
	
	The second factor corresponds to the shift in frequency caused by the change in oscillation position, and can be determined by differentiating equation \ref{eqn:Shifted Frequency} with respect to $d_\text{eq}$:
	\begin{align}
		\frac{\partial}{\partial d_\text{eq}}(f_\text{shifted}) = \frac{\partial}{\partial d_\text{eq}}\left(\sqrt{1 +  \frac{F_C'(d_\text{eq})}{k}}\frac{1}{2\pi}\sqrt{\frac{k}{m}}\right) \notag \\
		\implies \frac{\partial f_\text{shifted}}{\partial d_\text{eq}} = \frac{1}{4\pi\sqrt{km}}\frac{F_C''(d_\text{eq})}{\sqrt{1 + \frac{F_C'(d_\text{eq})}{k}}} \notag
	\end{align}
	
	The third factor corresponds to the change in oscillation position caused by the input force, and can be calculated by differentiating equation \ref{eqn:Equilibrium position definition} (rewritten in terms of $d_\text{eq} = x_C - x_\text{eq}$) with respect to $F_\text{in}$:
	\begin{align}
		\frac{\partial}{\partial F_\text{in}}\left(k(x_C - d_\text{eq})\right) = \frac{\partial}{\partial F_\text{in}}\left(F_C(d_\text{eq}) + F_\text{in} + F_\text{off}\right) \notag \\
		\implies \frac{\partial d_\text{eq}}{\partial F_\text{in}} = -\frac{1}{k}\frac{1}{(1 + \frac{F_C'(d_\text{eq})}{k})} \notag
	\end{align}
	
	Putting these together, the	final gain of the coupled oscillator framework is then given by
	\begin{align}
		G & \equiv \frac{1}{\Delta F_\text{in}} \notag \\
		& = -\frac{1}{4\pi}\frac{F_C''(d_\text{eq})}{\Delta f_\text{min}k^{3/2}m^{1/2}}\frac{1}{\left(1 + F_C'(d_\text{eq})/k\right)^{3/2}}
		\label{eqn:Gain}
	\end{align}
	
	This expression signals to us ways in which we can achieve high input force resolution by tweaking parameters of the system. The first way to increase gain is to decrease $\Delta f_\text{min}$ by being more sensitive to frequency shifts. Another way is to decrease $k$ or $m$ --- in this sense, coupled oscillators lend themselves to microelectromechanical systems (MEMS), since both mass and stiffness of materials decrease with system size \cite{MEMS textbook}.
	
\begin{figure}[!h]
		\includegraphics[width=\linewidth]{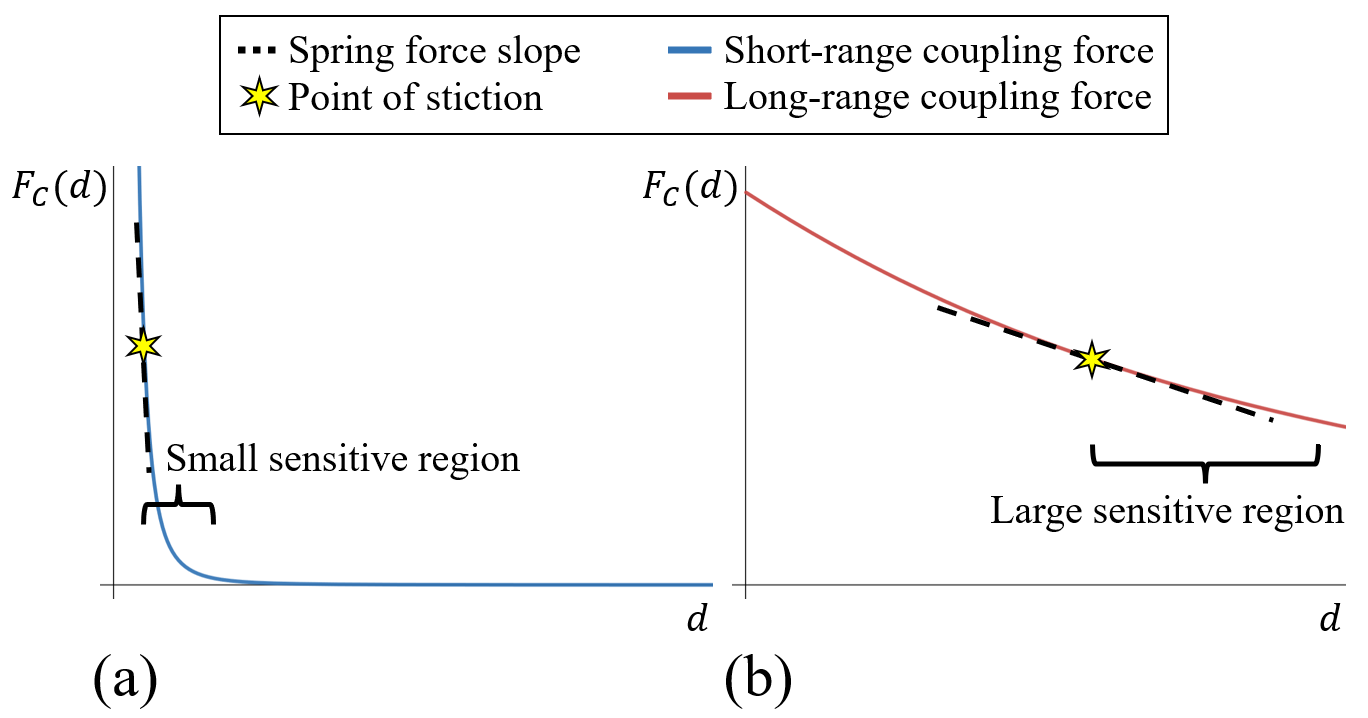}
		\caption{Comparison of different high-gain profiles for the coupling force. (a) Short-range force profiles can be sensitive where the curvature is high but will have a small operating region. (b) Long-range force profiles can be sensitive where the slope of the force is near the stiction value $-k$, which can provide a large operating region if the curvature is low.}
		\label{fig:Coupling force types}
	\end{figure}

	Finally, we can increase gain by finding a coupling force profile with certain properties. According to the expression for gain, there are two approaches to selecting coupling forces $F_C$ that minimize $\Delta F_\text{in}$ (fig. \ref{fig:Coupling force types}).
	
	The first is to increase the curvature $F_C''(d_\text{eq})$, while staying far away from the point of stiction (so $d_\text{eq} \gg d_\text{stic}$). This can be done by finding a short-range coupling force that diverges suddenly, so that the curvature is very high at the divergence. This approach lends itself to MEMS applications, because these naturally deal with short-range forces and oscillators will only displace by a small amount. For instance, the Casimir force between a sphere and a plate, which has a very sudden onset around $d \approx 100\text{nm}$, is one such short-range candidate that has been analyzed previously \cite{Josh's analysis paper}. MEMS systems that utilize or measure the Casimir force have been built previously (\cite{Zeptometer metrology paper}, \cite{Building Casimir on commercial MEMS paper}, \cite{Quantum actuation by Casimir paper}) and present excellent candidate platforms for a coupled oscillator system. One drawback of this approach is the increased difficulty of manufacturing due to the small distances necessarily involved: the oscillator must be placed at a very precise baseline distance from the coupling wall and there will be little spatial tolerance to avoid stiction. Such small systems are also more difficult to fabricate and assemble.
	
	The second method is to approach stiction and maximize the diverging term $(1 + F_C'(d_\text{eq})/k)^{-3/2}$. This requires that we have an attractive coupling force (so that $F_C' < 0$) and that we bring $F_C'(d_\text{eq})$ as close as possible to (while remaining lesser than) the negative spring constant $-k$. This does not universally mean having less room for error: proximity to stiction means $F_C'(d_\text{eq}) \gtrsim -k$ which does not necessarily restrict $x_\text{stic} - x_\text{eq}$ to be below any certain number. If $F_C''(d_\text{eq})$ is very low, $F_C'(d_\text{eq}) \gtrsim -k$ can be true over a large range of equilibrium distances $d_\text{eq}$. Further, since the term we are maximizing has a divergence, we can achieve extraordinary gain even with long-range forces. For these reasons, the system that we explore in the simulation is an example of a coupled oscillator that leverages a coupling force profile of this long-range type. One downside of this method is usually finding such a long-range interaction, and the fact that these forces tend to require large components as the coupling walls (e.g. large magnets or large metal plates) in order to have the desired properties, so the total size of the instrument increases. The oscillator itself, however, should remain as small as possible regardless of the chosen coupling force. Further, it can be difficult to find force profiles with steep enough curves so that they can overpower a spring and cause stiction at any point. Indeed, in the system we select, the required spring constant is of \SI{1.78e-3}{\newton/\meter}, which is much softer than even most MEMS springs.

\section{Coupled Oscillator Simulation}
	We programmed a discrete timestep simulation of the coupled oscillator system in MATLAB R2023b. There are 3 key variables that the simulation records over time: oscillator position $x(t)$, oscillator frequency $f(t)$, and the driving argument $\theta_D(t)$.
	
	The driving argument is used to calculate the driving force:
	\begin{equation*}
		F_D(t) \equiv A_D\cos(\theta_D(t))
	\end{equation*}
	where $A_D$ is a constant corresponding to the driving force amplitude.	We make $\theta_D$ a dynamic variable (i.e. updated in each timestep along with position and velocity) because the driving frequency is continuously changing, and we therefore need to update this driving argument in small timesteps to maintain continuity in the driving force.
	
	\subsection*{Main loop}
	In this section, we use square brackets (e.g. $x[t]$) to denote accessing of a discrete array rather than evaluating a continuous function. In each timestep of the simulation, we perform the following tasks:
	
	\begin{enumerate}
		\item \textbf{Record system state}. Save variable values $x[t]$ and $f[t]$ to memory.
		
		\item \textbf{Measure $f$}. First, we check if we are on a peak by looking at the positions recorded in the current and previous two steps ($x[t]$, $x[t - \Delta t]$, $x[t - 2\Delta t]$):
		\begin{align*}
			\ \ \ \ \ \ \ \  
			\left(x[t - \Delta t] > x[t - 2\Delta t]\right) \ & \text{AND} \ \left(x[t - \Delta t] > x[t]\right) \notag \\
			& \ \ \ 
			\implies \text{Peak at $t - \Delta t$}
		\end{align*}
		If we are on a peak, we use the time difference between the current peak and the $N_\text{cycles}$th previous peak to obtain an averaged estimate of the frequency of the oscillator. $N_\text{cycles}$ is the number of oscillation cycles we average over to better estimate the frequency:
		\begin{equation*}
			\ \ \ 
			f[t] = \frac{N_\text{cycles}}{t_\text{peaks}[n] - t_\text{peaks}[n - N_\text{cycles}]}
		\end{equation*}
		where $t_\text{peaks}$ is an array containing the times at which each peak occurs. The variable $n$ represents the current cumulative cycle count.
		
		Otherwise, if we do not detect a peak in this step, we retain the same frequency values as the previous step.
		
		One of the challenges of this implementation is frequency quantization: because of the discrete timestep, the period of the oscillation is resolvable only to limited precision, and therefore so is the frequency. For a timestep $\Delta t$ and frequency $f$, the frequency grain (smallest resolvable change in frequency) $\Delta f$ is
		\begin{align*}
			\Delta f & = \frac{1}{2N_\text{cycles}} \left(\frac{1}{1/f - \Delta t} - \frac{1}{1/f + \Delta t}\right) \\
			& \approx \frac{\Delta t f^2}{N_\text{cycles}}
		\end{align*}
		where we assumed that $\Delta t^2 \ll 1/f^2$ for the last expression. For low-frequency systems ($1-\SI{100}{\hertz}$) such as the one we study in the next section, this problem is only slightly noticeable, but when dealing with higher-frequency systems the frequency grain manifests as significant quantization of the measured frequency values. It should be noted that this is strictly a measurement problem and does not indicate that the simulation has underlying quantization: if we used a different method to measure frequency (e.g. Fourier transforms) this artifact would not exist.
		
		\item \textbf{Calculate the driving frequency $f_D$.} In order to enforce phase locking, the driving force can be in one of two states: matching or adjusting (fig. \ref{fig:Phase-locking states}).
		\begin{figure}[!h]
			\includegraphics[width=\linewidth]{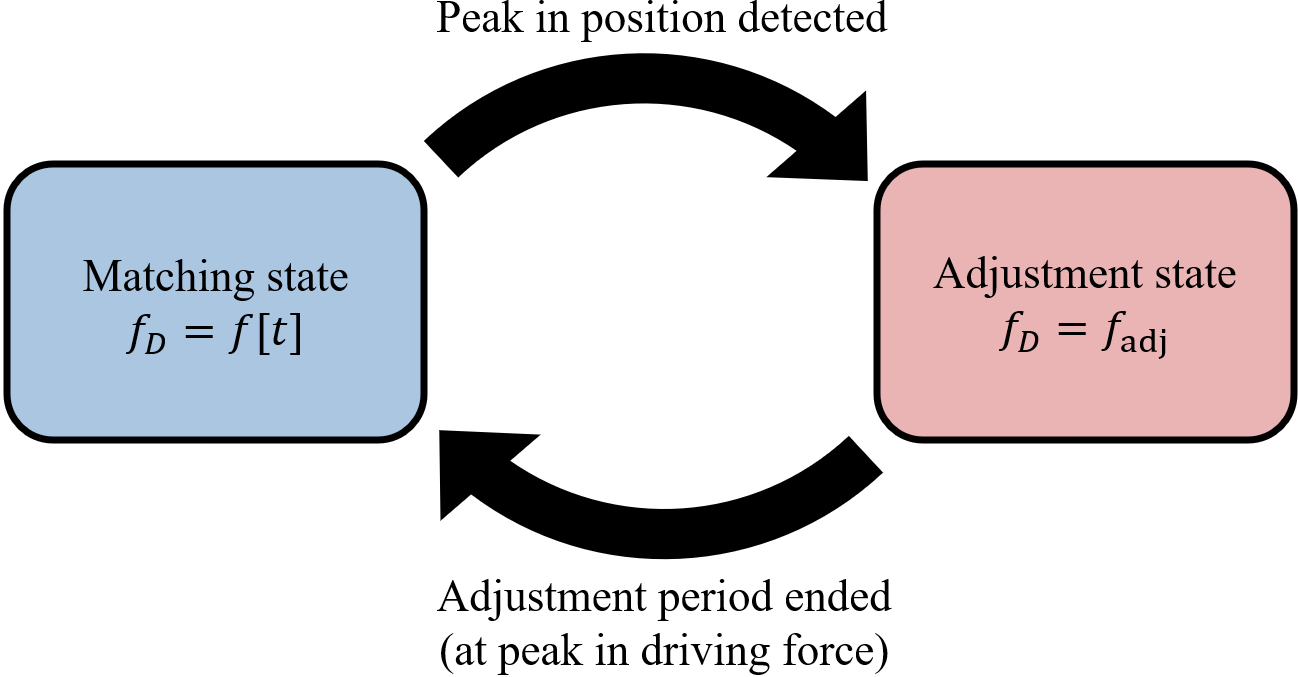}
			\caption{State diagram of the phase locking algorithm. When the oscillator reaches a peak in its motion, the program calculates $f_\text{adj}$, sets $f_D = f_\text{adj}$, and switches to the adjustment state. $f_\text{adj}$ is calculated such that the next peak in the driving force is at the correct place, given the current frequency of the oscillator $f[t]$ and the driving phase $\phi_D$. Once the adjustment period ends (at the next peak of the driving force curve), the program sets $f_D = f[t]$ and switches to the matching state.}
			\label{fig:Phase-locking states}
		\end{figure}		
		When the matching state is activated, the frequency of the driving force is set equal to the frequency of the oscillator at that time. The program will stay in this state and the driving frequency will remain unchanged until the oscillator reaches a peak in its motion, at which point it will set the driving frequency equal to the adjustment frequency $f_\text{adj}$ and switch to the adjustment state. $f_\text{adj}$ is calculated so that the next peak in the driving force occurs at the desired phase shift relative to the last peak in the motion of the oscillator
		\begin{equation}
			f_\text{adj} = \frac{\SI{360}{\degree} - \theta_D}{\phi_D}f[t] \notag
		\end{equation}
		where $\phi_D$ is the driving phase, which in our case is \SI{270}{\degree} (i.e. leading motion by \SI{90}{\degree}). In this way, we try to correct for any deviation from the desired phase by speeding up or slowing down the driving frequency and therefore enforce phase locking in the system. Once the system reaches the next peak in the driving force, we go back to the matching state (fig. \ref{fig:Phase-locking algorithm a}).
		\begin{figure}[!h]
			\includegraphics[width=\linewidth]{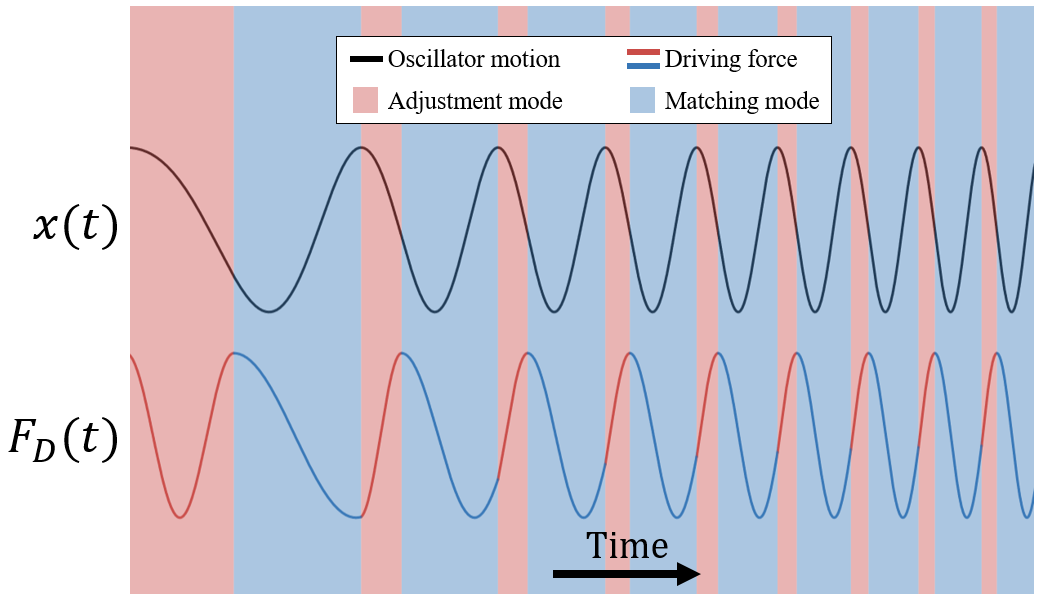}
			\caption{Example oscillator motion and how the driving force algorithm adjusts to maintain a phase of \SI{90}{\degree} behind the oscillator. Note that for the coupled oscillator framework, we require a driving phase that leads by \SI{90}{\degree}. Here we instead lag by \SI{90}{\degree} for visualization clarity, but the algorithm is the same for any phase.}
			\label{fig:Phase-locking algorithm a}
		\end{figure}
		This algorithm will approach the ideal driving force over time as long as the changes in frequency occur on a timescale much slower than the period of oscillation. The ideal driving force that it approaches would always have the same frequency as the motion, but would be shifted by exactly the desired phase at all times --- figure \ref{fig:Phase-locking algorithm b} shows how this algorithm produces a driving force that converges to the ideal for a slowly-varying frequency.
		\begin{figure}[!h]
			\includegraphics[width=\linewidth]{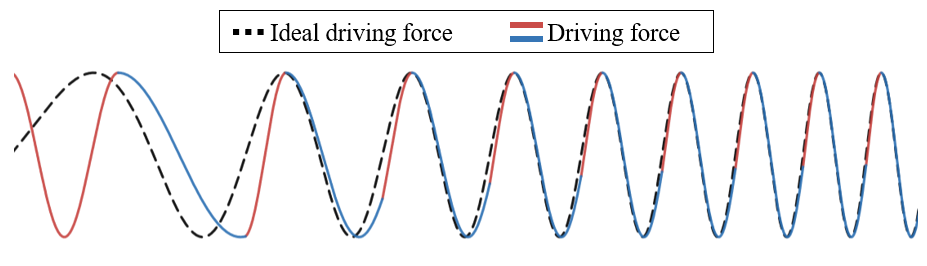}
			\caption{Driving force calculated from algorithm vs. ideal driving force. Over time, errors in phase are corrected and the two curves become the same.}
			\label{fig:Phase-locking algorithm b}
		\end{figure} 
		
		\item \textbf{Update the driving argument and get the driving force}:
		We use simple Euler integration to update the driving argument according to the current frequency, and get the driving force from it.
		\begin{align}
			\theta_D[t] = \theta_D[t - \Delta t] + f_D\Delta t \notag \\
			F_D = A_D\cos(\theta_D[t]) \notag
		\end{align}
			
		\item \textbf{Calculate the sum of forces and integrate the position and velocity of the system}:
		We use Newton's second law to calculate acceleration, and use Verlet integration to calculate the current velocity and position.
		\begin{equation*}
			\begin{array}{rl}
				\Sigma F = & -kx[t] - bv[t] \\
				& + F_D + F_C(x_C - x[t]) + F_\text{in} + F_\text{off} \\
				v[t] = & v[t - \Delta t] + \frac{1}{m}\Sigma F\Delta t \\
				x[t] = & x[t - \Delta t] + v[t]\Delta t
			\end{array}
		\end{equation*}
		
		\item \textbf{Check if stiction has occurred}:
		We check if the oscillator has experienced stiction, defined by
		\begin{equation}
			x[t] \geq x_C \implies \text{Stiction occurred} \notag
		\end{equation}
		If so, the simulation immediately terminates.
	\end{enumerate}
	
	\subsection*{Input force application}
\begin{figure}[!h]
		\includegraphics[width=\linewidth]{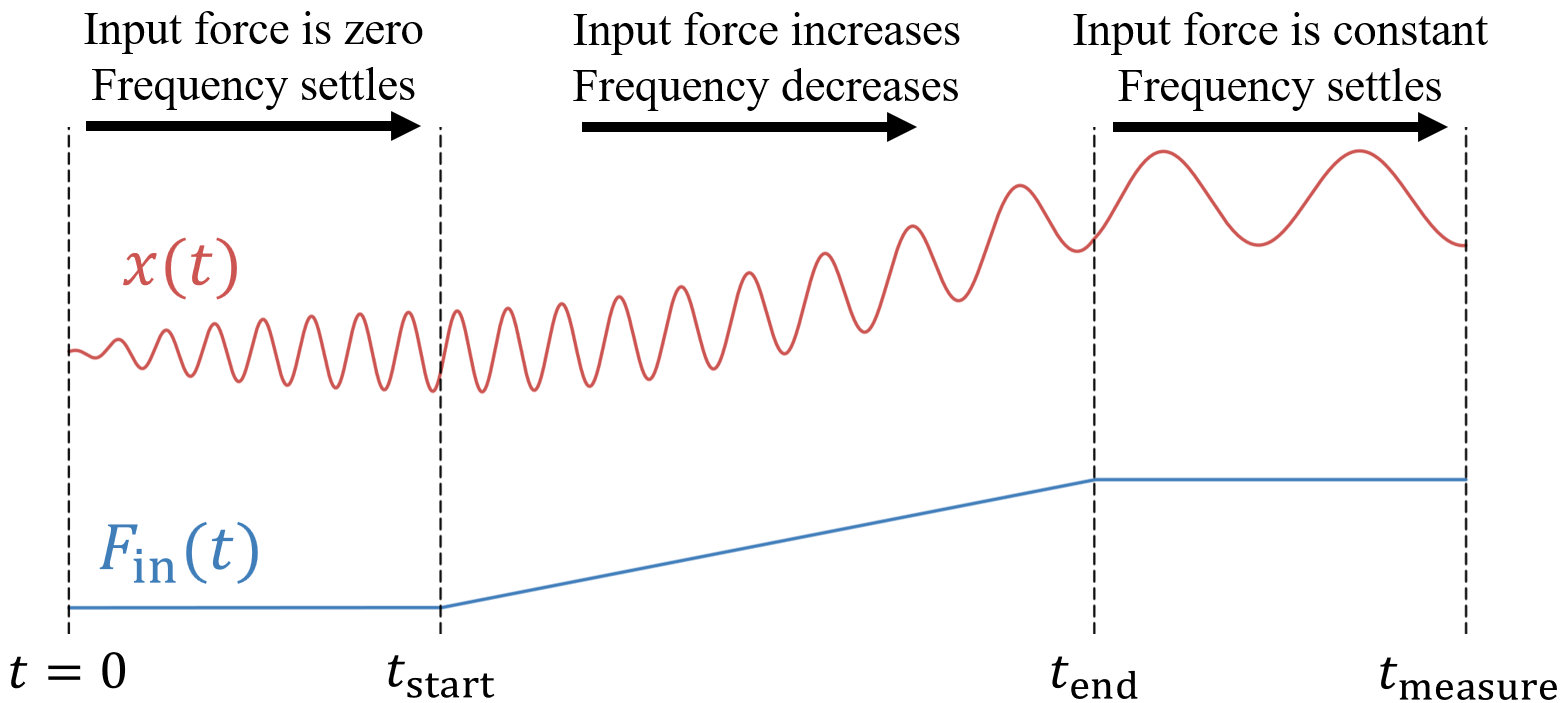}
		\caption{Input force onset. We slowly turn on the input force $F_\text{in}$ and then wait for the system to settle before we record the frequency. This graphic is for visual aid and is not a simulation output. The timescales selected here are for demonstration, but in the simulation there is much more time for transients to vanish compared to the oscillation period.}
		\label{fig:Input force onset}
	\end{figure}

	When the input force is applied, the exact equilibrium position of the oscillator (determined by equation \ref{eqn:Equilibrium position definition}) is generally unknown. Therefore, if we begin the simulation with an input force, the oscillator might initially be far from equilibrium and the ensuing high-amplitude transients may cross the point of no return $x_\text{no ret}$ and cause stiction. In order to avoid this issue, the system is always started with no input force so that the oscillator can be placed near the initial equilibrium position, which is well-known. After the transients have subsided, the input force begins to be increased at a constant rate, until the desired value is reached. The system is allowed to settle again before measurements are recorded. Although the input force is constant in the theoretical model, the system responds in a controlled manner to sufficiently slow changes in input force. Figure \ref{fig:Input force onset} illustrates the onset of the input force over time.
	
	There are then three important times in the simulation: the input force start time $t_\text{start}$, the input force end time $t_\text{end}$ and the measurement time $t_\text{measure}$. All should be much larger than the oscillation period to allow for a smooth transition into the input force application.
	
	\subsection*{Simulation parameters}
\begin{figure}[!h]
		\includegraphics[width=\linewidth]{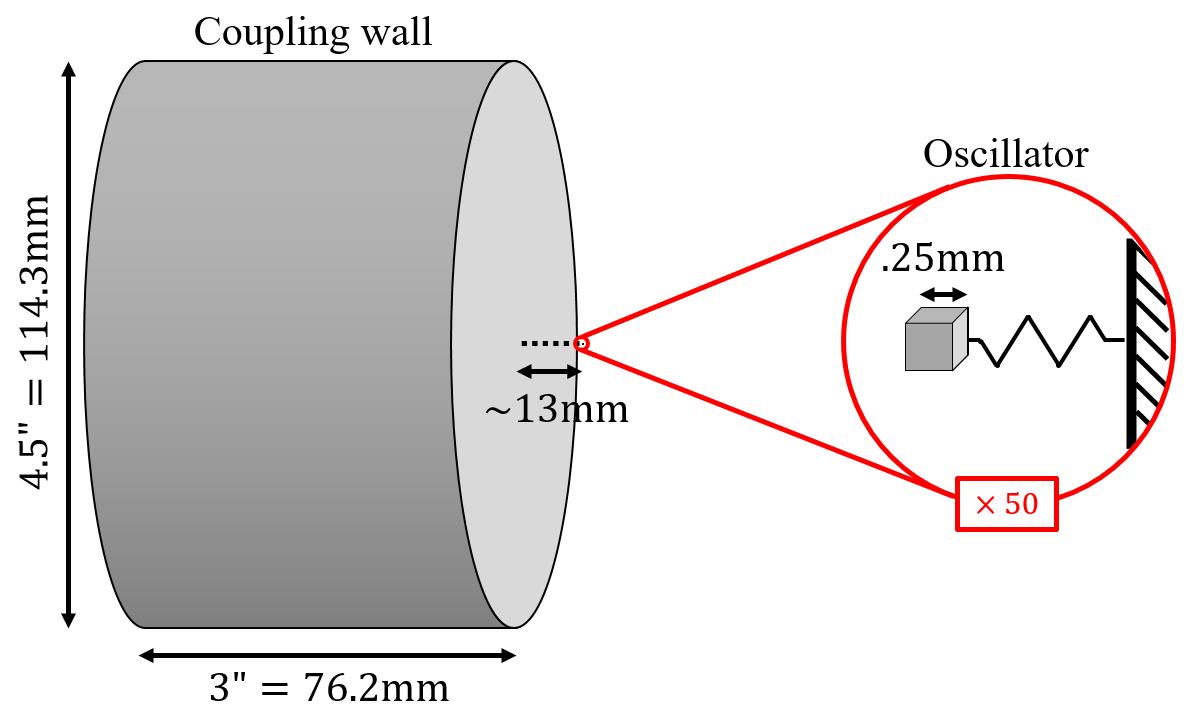}
		\caption{The coupled oscillator implemented in the simulation to demonstrate the framework. A large N52 magnet is used as the coupling wall, and a small one makes up the oscillator. Both have their magnetization directions aligned horizontally so that they experience attraction. The exact 
		distance between the magnets changes as an input force is applied, but it is always close to \SI{13}{\milli\meter}.}
		\label{fig:Simulated system}
	\end{figure}

	The system we simulated is outlined in figure \ref{fig:Simulated system}.
	Table \ref{tbl:Simulation parameters} contains detailed values of the parameters in this setup.
	\begin{table}[h!]
		\footnotesize
		\centering
		\caption{Parameter values in simulation}
		\label{tbl:Simulation parameters}
		\def\arraystretch{1.5}
		\begin{tabular}{|P{0.16\textwidth}|P{0.14\textwidth}|P{0.15\textwidth}|} 
			\hline
			
			\textbf{Parameter} & 
			\textbf{Value (SI units)} & 
			\textbf{Notes} \\
			\hline
			
			Timestep $\Delta t$ & 
			\SI{3e-4}{\second} & 
			Set by convergence tests \\
			\hline
			
			Input force start time $t_\text{start}$ & 
			\SI{25}{\second} & 
			Ample time for oscillations to settle \\
			\hline
			
			Input force end time $t_\text{end}$ & 
			\SI{75}{\second} & 
			Slow enough to avoid stiction due to transients \\
			\hline
			
			Measurement time $t_\text{measure}$ & 
			\SI{100}{\second} & 
			Ample time for oscillations to settle \\
			\hline
			
			Stiction distance $d_\text{stic}$& 
			\SI{12.7}{\milli \meter} & 
			Set to 0.5"\\
			\hline
			
			Starting distance $x_C$ & 
			\SI{13.335}{\milli \meter} & 
			$x_C = 1.05d_\text{stic}$\\
			\hline
			
			Mass $m$ & 
			\SI{11.875}{\micro \gram} & 
			Volume times density of \SI{250}{\micro \meter} cubical N52 magnet \cite{N52 magnet properties}\\
			\hline
			
			Spring stiffness $k$ & 
			\SI{1.78e-3}{\newton/\meter} & 
			$k = -F_C'(d_\text{stic})$\\
			\hline
			
			Damping coefficient $b$ & 
			\SI{4.5976e-9}{\kilogram/\second} & 
			Makes $Q = 1000$\\
			\hline
			
			Oscillation frequency at starting distance & 
			\SI{11.7}{\hertz} & 
			Uncoupled frequency of $f_0 = \SI{61.6}{\hertz}$\\
			\hline
			
			Driving force amplitude $A_D$ & 
			\SI{30}{\pico \newton} & 
			Ensures condition \ref{eqn:Simple linearity condition} \\
			\hline
			
			Offset force $F_\text{off}$& 
			\SI{-50.2}{\micro \newton} & 
			$F_\text{off} = -F_C(x_C)$ \\
			\hline
			
			Peaks averaged to get frequency $N_\text{cycles}$ & 
			20 & 
			Creates small frequency grain\\
			\hline
		\end{tabular}
	\end{table}
	\normalsize
	
	The oscillator is composed of a \SI{250}{\micro\meter}-side length cubical N52 magnet, which is the smallest commercially available N52-grade magnet, and is sold by SM Magnetics Co.. A 4.5"-diameter, 3"-thick cylindrical N52 magnet acts as the coupling wall. The only relevant detail of the coupling wall is the force it applies on the oscillator, which we numerically calculated using the Finite Element Method Magnetics software by building a model of the two-magnet system and sampling force data near the desired operating region. We then fit this data to a horizontally-shifted inverse-square function to obtain the coupling force:
	\begin{equation*}
		F_C(d)=\frac{\SI{1.72e-7}{\newton \meter^2}}{(d + \SI{4.52e-2}{\meter})^2}
	\end{equation*}

\section{Simulation Results and Analysis}
\begin{figure}[!h]
		\includegraphics[width=\linewidth]{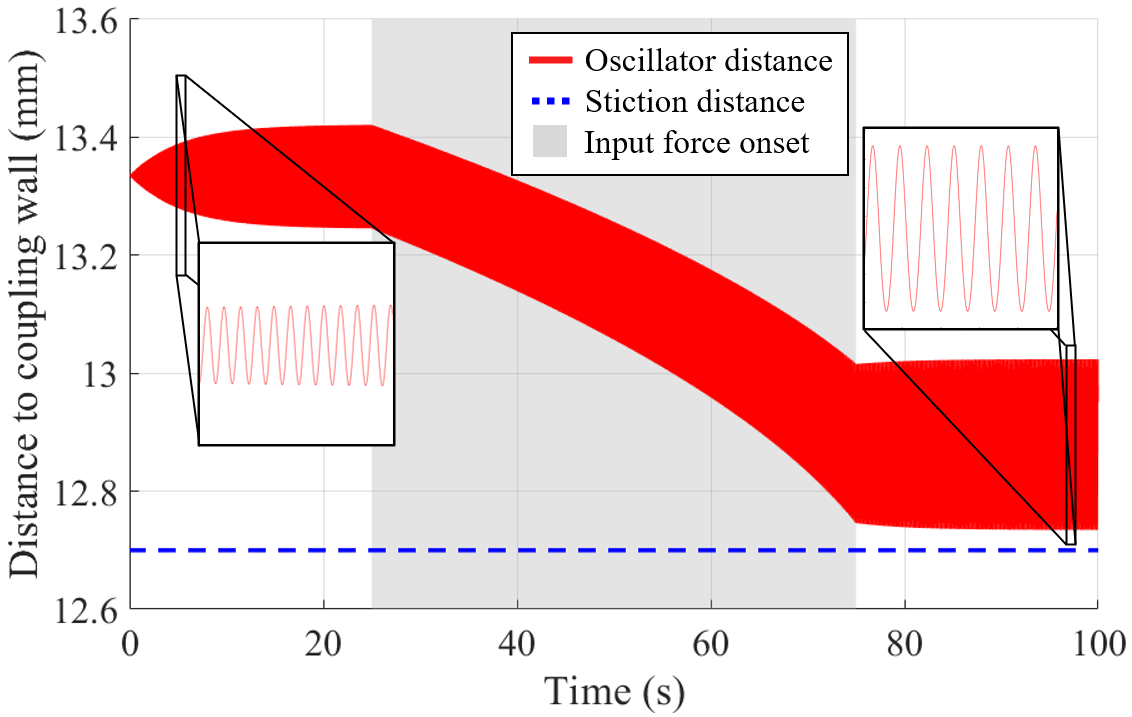}
		\caption{Distance from the oscillator to the coupling wall vs simulation time. The input force is gradually increased in the input force onset region, which moves the oscillator closer to the stiction distance and changes its oscillation frequency. The maximum input force applied is of \SI{20}{\nano \newton}. Note that by the end, the oscillation amplitude is comparable to the distance to stiction, so the linearity condition \ref{eqn:Simple linearity condition} is being violated.}
		\label{fig:Single run distance vs time}
	\end{figure}

	We ran the simulation with the settings outlined in table \ref{tbl:Simulation parameters}. Figure \ref{fig:Single run distance vs time} demonstrates a sample run with a final input force of \SI{20}{\nano \newton}. The oscillator starts at rest, with a driving frequency set near resonance. After a few oscillations, phase-locking is engaged and the driving frequency quickly shifts to the exact undamped resonance frequency. At $t_\text{start} = \SI{25}{\second}$, the input force begins to be applied. The input force continues increasing linearly until $t_\text{end} = \SI{75}{\second}$, at which point the input force stays at a fixed value and the oscillator is allowed to settle again before the simulation ends at $t_\text{measure} = \SI{100}{\second}$. At the end of the simulation, the frequency has been significantly decreased by an amount that corresponds to the input force applied.
	
\begin{figure}[!h]
		\includegraphics[width=\linewidth]{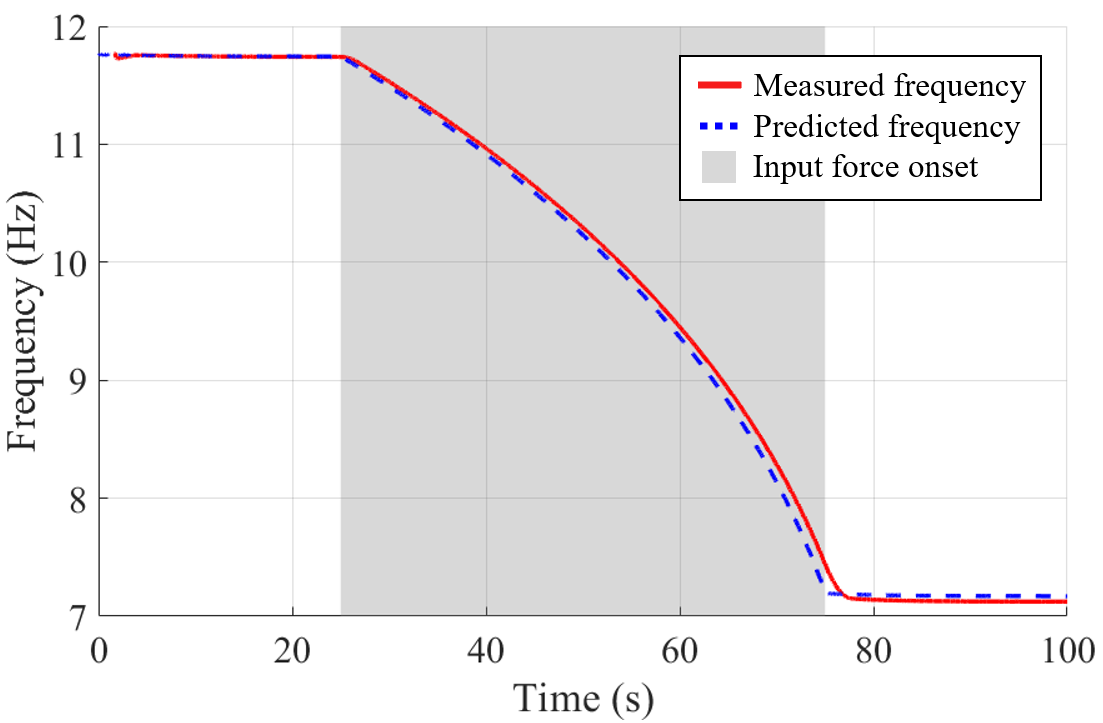}
		\caption{Measured and predicted frequencies of the oscillator vs time in a single run of the simulation. Note that there is a brief time at the start where there have not been enough periods of oscillation to take the average of, so there is no measured frequency. The predicted frequency is calculated using equation \ref{eqn:Shifted Frequency} with the equilibrium distance of the oscillator measured in the simulation as an input.}
		\label{fig:Single run frequency vs time}
	\end{figure}

	Figure \ref{fig:Single run frequency vs time} shows the measured and predicted frequencies (equation \ref{eqn:Shifted Frequency}) as the system evolves in time. For the latter, the equilibrium distance is used as an input, which is calculated by averaging the positions of the last two extrema. At $t = \SI{20}{\second}$, the predicted frequency is within $0.1\%$ of the measured value. At $t = \SI{95}{\second}$, it is within $1\%$. The decreased quality of the prediction can be attributed to the violation of the linearity condition \ref{eqn:Simple linearity condition}, which was held to derive the expressions.
	
\begin{figure}[!h]
		\includegraphics[width=\linewidth]{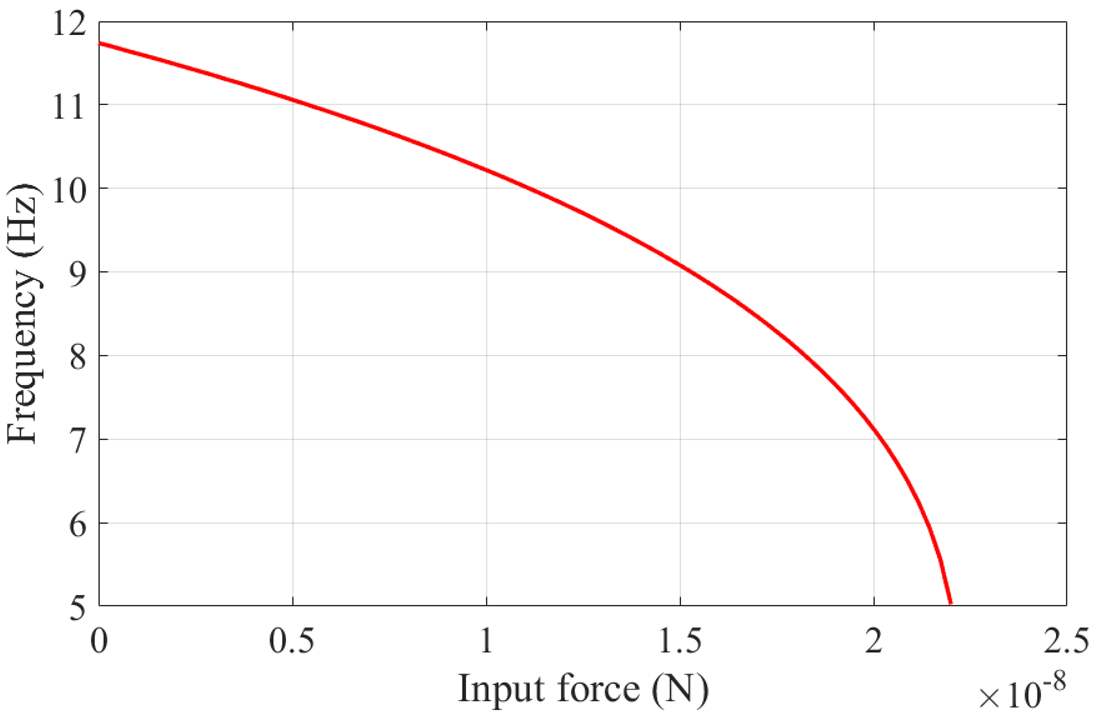}
		\caption{Final oscillator frequency vs final input force over multiple simulation runs. Because we are near stiction, this relationship is highly nonlinear, and tends towards a vertical asymptote as stiction is reached.}
		\label{fig:Frequency vs input force}
	\end{figure}

	We then ran the simulation multiple times with final input forces ranging from \SI{0}{\nano \newton} to \SI{22}{\nano \newton}. Figure \ref{fig:Frequency vs input force} shows the final frequency vs final input force of the system. The frequency decreases non-linearly as the input force causes the system to approach stiction, and would reach zero at the point of stiction. 
	
\begin{figure}[!h]
		\includegraphics[width=\linewidth]{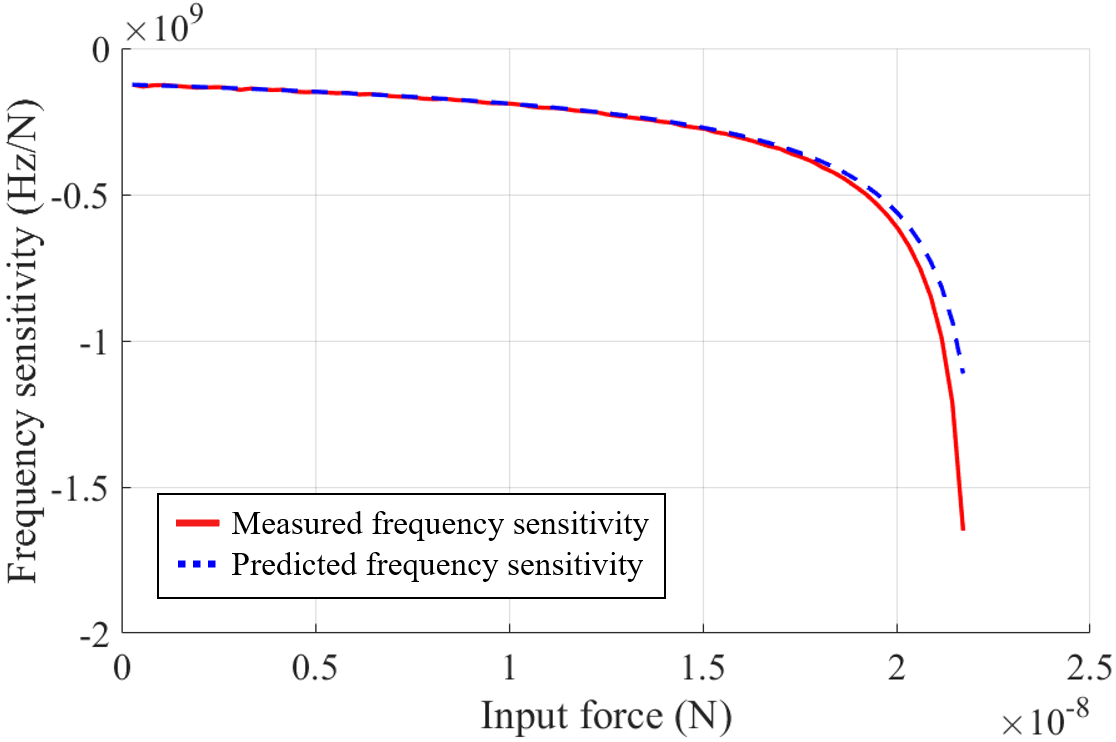}
		\caption{Measured and predicted frequency sensitivities. The prediction is made using equation \ref{eqn:Frequency sensitivity} and the final equilibrium distance of the oscillator.}
		\label{fig:Frequency sensitivity vs input force}
	\end{figure}

	For the purposes of measurement sensitivity, we are interested in how rapidly the frequency changes per input force increment. We can study this by taking the derivative of the data in figure \ref{fig:Frequency vs input force}, which is shown in figure \ref{fig:Frequency sensitivity vs input force}. We overlay the theoretical frequency sensitivity $\left(\frac{\partial f_\text{shifted}}{\partial F_\text{in}}\right)_\text{theory}$, calculated using the gain expression \ref{eqn:Gain}:
	\begin{align}
		\left(\frac{\partial f_\text{shifted}}{\partial F_\text{in}}\right)_\text{theory} & = \Delta f_\text{min} G \notag \\
		& = -\frac{1}{4\pi}\frac{F_C''(d_\text{eq})}{k^{3/2}m^{1/2}}\frac{1}{\left(1 + F_C'(d_\text{eq})/k\right)^{3/2}}
		\label{eqn:Frequency sensitivity}
	\end{align}
	We multiply by $\Delta f_\text{min}$ because the simulation does not include the final step of measuring the frequency.
	
	The gain expression matches the simulated output within $3\%$ for $F_\text{in} = \SI{5}{\nano\newton}$, and within $20\%$ for $F_\text{in} = \SI{21}{\nano\newton}$. Once again, proximity to stiction violates the assumptions on which we built this expression, so this is not surprising. We note that, due to frequency quantization, frequency sensitivity appears noisy for small input forces. This effect is visible as small ripples in the low-input force region. For this reason, the 3\% figure was calculated after smoothing the data with a 5-point square-window moving average.
	
\section{Analysis and Application to Gradiometry}
	\subsection*{Identification of Frequency Grain}
	With this system we achieved simulated sensitivities of $\sim$\SI{5e8}{\hertz/\newton}. In order to convert this to a force resolution, we must know the smallest frequency shift we can measure $\Delta f_\text{min}$. This number depends on the quality of the instrumentation used to make the measurement, the noise characteristics of the system, and tuned parameters like the averaging/gate time of measurement.
	
\begin{figure}[!h]
		\includegraphics[width=\linewidth]{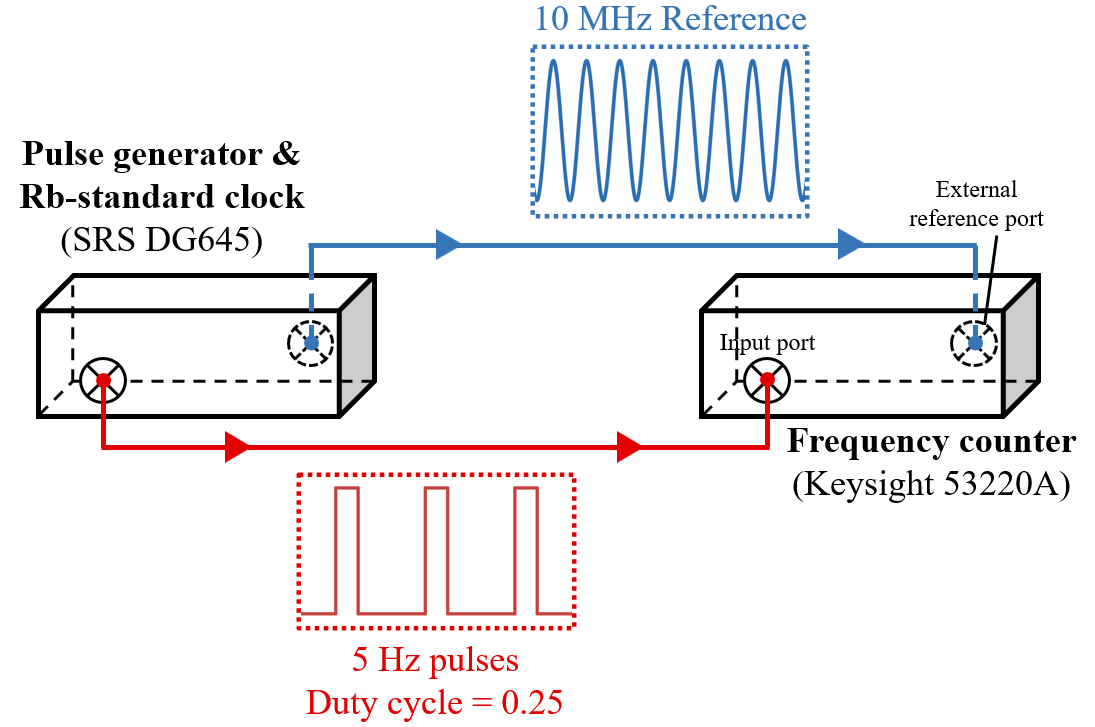}
		\caption{System used to determine the frequency resolution. The DG645 provides a rubidium-standard reference signal to the 53220A frequency counter to improve resolution. The same DG645 is used to generate 5 Hz pulses with a duty cycle of 0.25 which act as the test signal.}
		\label{fig:Frequency resolution testing system}
	\end{figure}

	To find an adequate value for this number, we fed a \SI{5}{\hertz} square pulse wave with a .25 duty cycle to a Keysight 53220A frequency counter. The input wave was generated using a Stanford Research Systems DG645 Digital Delay/Pulse Generator. To further increase frequency resolution, we used the \SI{10}{\mega\hertz} reference signal from the DG645 as an external timebase reference for the frequency counter. All signals coming from the DG645 are rubidium-standard. Figure \ref{fig:Frequency resolution testing system} shows this system.
		
\begin{figure}[!h]
		\includegraphics[width=\linewidth]{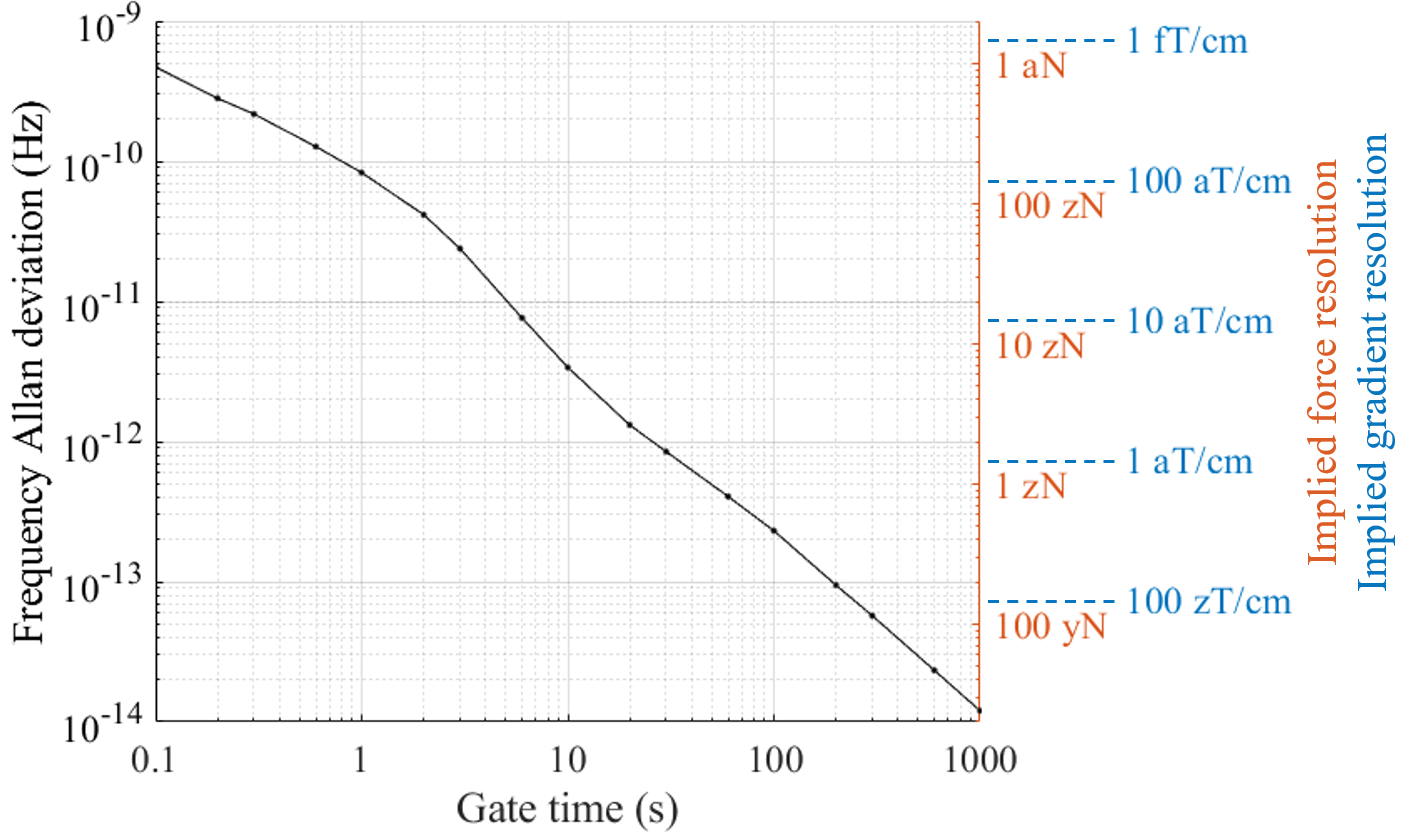}
		\caption{Allan deviation (frequency resolution) vs gate time for measurement. As the gate time increases, our resolution improves due to a longer averaging time. On the right axis, we show the implied force resolution (red) and gradient resolution (blue). The force resolution can be found by dividing the frequency resolution by the sensitivity of the coupled oscillator (\SI{5e8}{\hertz/\newton}) and the gradient resolution can be found by dividing the force resolution by the magnetic moment of the oscillator (\SI{1.5e-3}{\tesla/\centi\meter})}
		\label{fig:Allan deviation plot}
	\end{figure}

	Using this system, we average 50 measurements (where one measurement corresponds to the completion of one gate cycle) and obtain the Allan variance of the measured frequency for a range of gate times using the built-in statistics feature of the frequency counter. The Allan variance serves as a measure of the frequency resolution at that gate time. See figure \ref{fig:Allan deviation plot} for this relationship.
		
	Note that we cannot observe any ``bottoming out'' of the Allan deviation due to frequency drift as we went to higher gate times, as is generally expected of such a study. \SI{1000}{\second} is the maximum gate time achievable on the Keysight 53220A, so if such drift effects exist, we cannot detect them with this setup. Regardless, due to the shared reference signal between the source and counter, we expect any contributions from drift to be exceedingly low.
	
	From this study, we see that at \SI{5}{\hertz} and using a \SI{1}{\second} gate time, the measurement system described above is capable of resolving frequency shifts of just below $\Delta f_\text{min} = \SI{e-10}{\hertz}$.

	\subsection*{Force resolution}	
	Combined with the coupled oscillator sensitivity of \SI{5e8}{\hertz/\newton}, our frequency resolution of $\Delta f_\text{min} = \SI{e-10}{\hertz}$ would imply a force resolution of $\Delta F_\text{min} = \SI{200}{\zepto\newton}$. If instead we use a \SI{100}{\second} gate time, the frequency resolution becomes $\Delta f_\text{min} = \SI{2.3e-13}{\hertz}$, implying a force resolution of $\Delta F_\text{min} = \SI{460}{\yocto\newton}$. See figure \ref{fig:Allan deviation plot} for a more detailed view of force resolution versus gate time.
	
	\subsection*{Application to magnetic gradiometry}
	We can apply this system to the measurement of magnetic gradients readily. Because the oscillator is itself a magnet, magnetic gradients in the oscillation direction will induce a force that we can treat as the input force. The system is therefore sensitive to magnetic gradients at a single point (within the oscillator magnet dimensions of \SI{250}{\micro\meter}). The oscillator magnet has a magnetic moment \SI{1.5e-5}{\newton/(\tesla/\meter)}, so the smallest detectable gradient is of $\SI{13}{\femto\tesla/\meter} = \SI{130}{\atto\tesla/\centi\meter}$ at a \SI{1}{\second} gate time and $\SI{31}{\atto\tesla/\meter} = \SI{310}{\zepto\tesla/\centi\meter}$ at a \SI{100}{\second} gate time.
	
	Previous instances of \SI{}{\atto\tesla/\centi\meter} coupled oscillator gradiometry have been described \cite{Josh's analysis paper}. However, these have had \SI{}{\nano\meter}-size constraints due to a reliance on the Casimir effect as a coupling force, which is exceedingly short-range. This work presents comparable sensitivity using a micro-scale rather than nano-scale construction.

	\subsection*{Theory to practice}
	We have described and simulated an example of coupled oscillators that is based on realistic magnetic force curves. The technology and micro-assembly techniques required to build this device are standard MEMS processes and have been demonstrated extensively in the past (e.g. \cite{Building Casimir on commercial MEMS paper}, \cite{Quantum actuation by Casimir paper}, \cite{100pT/cm paper}) and we believe that the next step is construction of the device. To this end, there are technical obstacles that must be overcome and whose effects may detriment the sensitivity of a realization of the proposed device. We foresee the most difficult of these to be the following:
	
	\begin{enumerate}
		\item \textbf{Noise:}
		The system we described assumes a noiseless environment. However, in any laboratory implementation of this device, there are several sources of noise that limit sensitivity if not sufficiently suppressed \cite{MEMS textbook}. Thermal and mechanical noise may require cooling of the system and geometric design that avoids unwanted resonances. An analysis of electrical noise requires defining how the position of the oscillator is measured and converted into an electrical signal. Particularly, the low-frequency nature of the system may make it more vulnerable to $1/f$ fractional noise sources. Phase-lock amplifiers and emerging artificial intelligence noise-reduction schemes \cite{Noise reduction paper} would help lower the electrical noise significantly --- however, care should be taken in adding steps to the signal pipeline as they may create a bottleneck in frequency precision and ultimately hinder sensitivity. Finally, in the above implementation using magnets, fluctuations in the gradient of the Earth's magnetic field may be a significant source of noise as we are operating close to and below the geomagnetic gradient noise floor of \SI{500}{\femto\tesla/\cm} \cite{Josh's analysis paper} (regardless of whether the system is being applied to gradiometry). Overcoming this source of noise would require magnetic shielding of the apparatus, but it should be noted that even without shielding, this theoretical framework can detect much smaller signals than competing technologies that measure magnetic fields directly, since the noise floor for geomagnetic fields is much larger than for geomagnetic field gradients, compared to sources of interest (such as the heart).
		
		\item \textbf{Damping:}
		In order to achieve a quality factor of $Q = 1000$, the oscillator will likely need to be placed in vacuum, especially since the spring constant and mass are particularly low. It may be possible to build this device with a lower quality factor and retain the desired sensitivity, but such analysis is beyond the scope of this work.
		
		\item \textbf{Weakness of driving force:}
		The driving force was of \SI{30}{\pico\newton} in amplitude, which may be difficult to achieve with traditional driving mechanisms such as parallel electrodes. Once again, this is in large part because the spring constant is so small that the system is easily excitable, and in order to maintain small oscillation amplitudes so close to stiction we require a small driving force.
		
		\item \textbf{Softness of spring:}
		The spring in the proposed model has a low stiffness of $k = \SI{1.78e-3}{\newton/\meter}$, which is difficult to fabricate without increasing susceptibility to undesired vibrational modes. Particularly, long spring or cantilever designs would be sensitive to torsion, which would make the oscillator sensitive to magnetic fields and introduce high noise floors associated with them.
		
		\item \textbf{Reduced dynamic range due to stiction:}
		A strong enough input force has the potential to induce stiction. In the system described, an input force of \SI{23}{\nano\newton} would be sufficient to cause stiction. This may be avoided by making the offset force dynamic using a proportional-integral-derivative (PID) controller, so that it matches the coupling force even as the system is offset from the origin.
	\end{enumerate}
	
	Three of these points (damping, weakness of driving force, and softness of spring), can be aided by increasing the coupling force strength or using a larger magnet as the oscillator. This would lead to a stiffer force potential and therefore a stiffer spring which would demand less damping and a stronger driving force. Despite these potential difficulties, the system described above is made of simple parts and shows a potential gradient sensitivity that is well in excess of the top gradiometry approaches \cite{100pT/cm paper}.

\section{Acknowledgments}
We would like to recognize the great value of discussions with Nicholas Fuhr and Zhancheng (Ryan) Yao's during the making and tuning of this simulation. This work was supported in part by the SONY Chip-scale Magnetocardiography grant and Boston University.

\clearpage
\section*{Appendix: Tracking Resonant Frequency via Phase-Locking in a Driven Oscillator}
Frameworks such as coupled oscillators transduce a measurement (for instance, of a force) into a resonant frequency shift in an oscillator. That is, the measured quantity changes the parameters of the oscillator in such a way that the resonant frequency is affected. However, in order to detect these shifts we need a mechanism to make the actual oscillation frequency track this change in resonance. In particular, a driven, damped oscillator in steady state will only oscillate at exactly the frequency at which it is driven, so if our driving frequency never changes, our oscillation will stay at the same frequency regardless of what the resonance frequency is and we will not be able to track our signal.

Therefore, a feedback system that changes the driving frequency to match the change in resonance is required. In the coupled oscillator framework, this mechanism is the phase-locking of the driving force to be $\SI{90}{\degree}$ ahead of the motion of the oscillator. In what follows, we will present a mathematical model of phase locking and show that it leads to tracking of the resonant frequency. More precisely, we will show that this system tracks the undamped resonance frequency, regardless of what the value of damping is.

\subsection*{Phase-locking model and fixed point}
Our system consists of a damped harmonic oscillator of mass $m$, damping constant $b$ and spring constant $k$, with a phase-locked driving force. The complex equation of motion of this system can be written as:
\begin{equation}
	\ddot x(t) + \gamma\dot{x}(t) + \omega_0^2 x(t) = f_D(x(t))
	\label{Equation of motion in x}
	\tag{A1}
\end{equation}
where $x(t)\in\mathbb{C}$ is the complex position of the oscillator (with the physical position being $\operatorname{Re}(x)$), $\gamma \equiv b/m$ is the damping constant and $\omega_0 \equiv \sqrt{\frac{k}{m}}$ is the undamped resonance frequency. Note that in this appendix, we work in frequency units of $\SI{}{\radian/\second}$, whereas in the main text of this paper we worked in units of rev/$\SI{}{\second}$, or $\SI{}{\hertz}$. $F_D(x)\in\mathbb{C}$ is the phase-locked driving force (with physical driving force $\operatorname{Re}(F_D(x))$), and $f_D(x) \equiv \frac{1}{m}F_D(x)$ is the reduced phase-locked driving force. Note that the driving force depends on the position of the oscillator due to phase locking: the force is always at a fixed complex phase relative to the oscillator.

In order to describe the phase locking behavior of the driving force, we will first rewrite our complex position in radial form without loss of generality
\begin{equation*}
	x(t) = A(t)e^{i\theta(t)}
\end{equation*}
Where we call $A(t)\in\mathbb{R}$ the instantaneous oscillation amplitude and $\theta(t)\in\mathbb{R}$ the instantaneous oscillation angle. Our phase-locked force can then be written as
\begin{equation*}
	F_D(x) = F_D(\theta) = F_{D,0}e^{i(\theta(t) - \phi_D)}
\end{equation*}
Where $F_{D,0}\in\mathbb{R}$ is the driving amplitude and $\phi_D\in\mathbb{R}$ is the driving phase, both of which are constant. Notice that the oscillations in this force necessarily lag the oscillator in phase by $\phi_D$ at all times -- this is how we represent phase locking analytically.

By plugging in $x=Ae^{i\theta}$ and $F_D(x) = F_{D,0}e^{i(\theta - \phi_D)}$ into equation \ref{Equation of motion in x} and separating the real and imaginary parts, we get new equations of motion for $A$ and $\theta$:
\begin{align*}
	\ddot A + \gamma \dot A + (\omega_0^2 - \dot\theta^2)A &= f_{D,0}\cos(\phi_D)   \\
	A \ddot\theta + (\gamma A + 2\dot A) \dot\theta &= -f_{D,0}\sin(\phi_D)
\end{align*}
Since only derivatives of $\theta$ show up in these equations, we write in terms of the instantaneous frequency of oscillation $\omega(t) \equiv \dot\theta(t)$:
\begin{align}
	\ddot A + \gamma \dot A + (\omega_0^2 - \omega^2)A &= f_{D,0}\cos(\phi_D) \notag\\
	A \dot\omega + (\gamma A + 2\dot A) \omega &= -f_{D,0}\sin(\phi_D)
	\label{eqn:Equations of motion}
	\tag{A2}
\end{align}

By setting $\ddot A,\dot A,\dot \omega = 0$ we find a fixed-point of this system $(\omega_f, A_f)$: 
\begin{align}
	\begin{split}
		\omega_f &= \omega_0 \left(\sqrt{1 + \left(\frac{\gamma}{2\omega_0\tan(\phi_D)}\right)^2} + \frac{\gamma}{2\omega_0\tan(\phi_D)}\right)\\
		A_f &= -\frac{f_{D,0}\sin(\phi_D)}{\omega_0\gamma} \\
		&\left(\sqrt{1 + \left(\frac{\gamma}{2\omega_0\tan(\phi_D)}\right)^2} - \frac{\gamma}{2\omega_0\tan(\phi_D)}\right) \label{eqn:Steady-state amplitude and frequency}
	\end{split}
	\tag{A3}
\end{align}

Our ultimate goal is to show that the dynamical system described always goes towards this fixed point, which we will show in the following section. With this, we would show that phase locking adequately tracks changes in oscillation frequency. Note that, for the phase used in the coupled oscillator framework of $\phi_D = 3\pi/2$, $\omega_f = \omega_0$. Therefore, if this fixed point represents the steady state of the system, then phase locking will track the undamped oscillation frequency $\omega_0$.

Note also that the fixed points are negative for $0 < \phi_D < \pi$. In the simulation and in a real-life phase-locked loop, this corresponds to deamplification. The simulation breaks for these phase values, because the oscillation speed increases without bound until the period of oscillation is below the timestep of the simulation. For this reason, we will focus our analysis on the amplification region, with $\pi < \phi_D < 2\pi$.

\subsection*{Convergence of fixed point}
\begin{figure}[!h]
	\includegraphics[width = \linewidth]{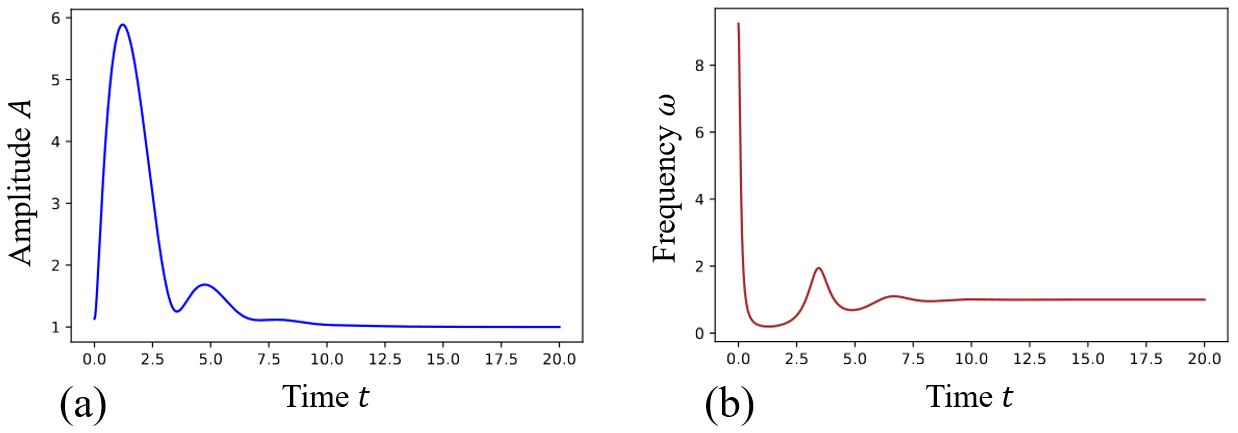}
	\caption{Numerically calculated amplitude (a) and frequency (b) of the oscillator over time at a fixed phase of $3\pi/2$.}
	\label{fig:Amplitude and frequency vs time}
\end{figure}

We will show evidence of the stability of the fixed point by numerically solving the equations and observing convergence to the fixed points, and then by showing that the fixed point is linearly stable. By numerically solving equations \ref{eqn:Equations of motion}, we find $A(t)$ and $\omega(t)$ at $\phi = 3\pi/2$ (fig. \ref{fig:Amplitude and frequency vs time}). In this solution, we set $m, b, k, F_{D,0} = 1$, so equations \ref{eqn:Steady-state amplitude and frequency} imply that $\omega_f = 1$ and $A_f = 1$. Indeed, this is the behavior shown in figure \ref{fig:Amplitude and frequency vs time}. 

\begin{figure}[!h]
	\includegraphics[width=\linewidth]{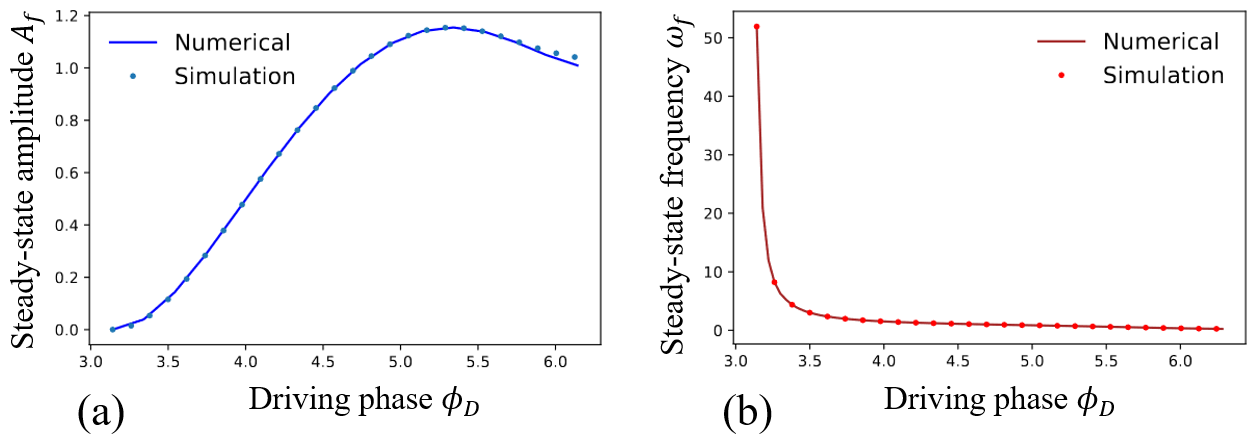}
	\caption{Steady-state amplitude (a) and steady-state frequency (b) of the oscillator vs driving force phase. The numerical curve is obtained by solving equations \ref{eqn:Equations of motion} and the simulation data is obtained from the algorithm described in the main text. Both of these closely match the curves predicted in equations \ref{eqn:Steady-state amplitude and frequency}.}
	\label{Ampfreqvsphase}
\end{figure}

In figure \ref{Ampfreqvsphase}, we vary $\phi_D$ between $\pi$ and $2\pi$ and observe the steady-state values of $A$ and $\omega$. These are calculated as $A_f = A(t = 20)$ and $\omega_f = \omega(t = 20)$. The values in the simulation match closely with the steady-state values produced by directly solving equations \ref{eqn:Equations of motion} as in figure \ref{fig:Amplitude and frequency vs time}.

The trends of convergence towards predictable steady states seen in the plots have held unconditionally for every set of parameters that were tested -- however this does not prove that the system will always converge. To show the convergence of the amplitude and the frequency of the oscillator analytically, we rewrote equations \ref{eqn:Equations of motion} as three first order equations by introducing a variable $V \equiv \dot A$:
\begin{align}
	\begin{split}
		\dot \omega &= -\left(\gamma + \frac{2V}{A}\right)\omega - \frac{f_{D,0} \sin{\phi_D}}{A}\\
		\dot A &= V\\
		\dot V &= -\gamma V - \left(\omega_0^2 - \omega^2\right)A+f_{D,0} \cos{\phi_D}
	\end{split}
	\tag{A4}
\end{align}

The fixed point given in equation \ref{eqn:Steady-state amplitude and frequency} can be recovered by setting $\dot \omega, \dot V, \dot A = 0$, with the additional coordinate $V_f = 0$. To study the stability of the fixed point, we linearize the above equation by a change of variables $\Delta\omega \equiv \omega - \omega_f$, $\Delta A \equiv A - A_f$ and $\Delta V \equiv V - V_f = V$. We also write $-f_D\sin{\phi_D} = \gamma \omega_f A_f$ and $f_D\cos{\phi_D} = (\omega_0^2 - \omega_f^2)A_f$, which can be shown from the definition of the fixed point. The equations become:

\begin{align*}
	\begin{split}
		\dot{\Delta\omega} &= -\gamma \Delta\omega - \gamma \frac{\omega_f}{A_f}\Delta A - 2\frac{\omega_f}{A_f}\Delta V\\
		\dot{\Delta A} &= \Delta V\\
		\dot{\Delta V} &= 2\omega_f A_f \Delta\omega - (\omega_0^2 - \omega_f^2)\Delta A - \gamma\Delta V
	\end{split}
\end{align*}

The Jacobian $J$ is then:
\begin{align*}
	\begin{pmatrix} 
		-\gamma & -\gamma\frac{\omega_f}{A_f} & -2\frac{\omega_f}{A_f}  \\  
		0 & 0 & 1 \\
		2A_f \omega_f & -(\omega_0^2-\omega_f^2) & -\gamma
	\end{pmatrix}
\end{align*}
We can then look at the sign of the real part of the eigenvalues of this matrix to determine the linear stability of the fixed point $(\omega_f,A_f, V_f)$. The eigenvalues satisfy the following polynomial expression:
\begin{equation}
	\lambda^3+2\gamma\lambda^2+(\gamma^2+\omega_0^2+3\omega_f^2)\lambda+(\omega_f^2+\omega_0^2)\gamma = 0
	\label{Eigenvalue polynomial}
	\tag{A5}
\end{equation}

Since all the coefficients are real and positive, the cubic polynomial has one negative real root $\lambda_0$ and two complex roots $\lambda_+, \lambda_-$ that are each other's complex conjugates ($\lambda_+ = \lambda_-^*$).

We know that
\begin{equation*}
	\tr{(J)} = \lambda_0 + \lambda_+ + \lambda_-
\end{equation*}
So we find that
\begin{equation*}
	-\gamma - \lambda_0/2 = \operatorname{Re}(\lambda_+) = \operatorname{Re}(\lambda_-)
\end{equation*}
Plugging $\lambda = -2\gamma$ into the polynomial in equation \ref{Eigenvalue polynomial} gives a negative value, and since the coefficients are all positive, it must be true that $\lambda_0 > -2\gamma$. Therefore, we find that
\begin{equation*}
	-\gamma - (-2\gamma)/2 > \operatorname{Re}(\lambda_+) = \operatorname{Re}(\lambda_-)
\end{equation*}
So
\begin{equation*}
	\operatorname{Re}(\lambda_+) = \operatorname{Re}(\lambda_-) < 0
\end{equation*}

Therefore, all our eigenvalues have negative real parts and the fixed point exhibits linear stability. While this is not a complete proof of the convergence of this system toward the fixed point (as we have not shown that there are no cycles or divergences in the flow), the linear stability and the evidence from solving the differential equations and running the associated simulation point to total convergence. All the initial conditions and parameters tested during this project have exhibited rapid convergence to the fixed point, and we believe this to be the case with any possible initial conditions and parameters given that $\pi<\theta_D<2\pi$.
\end{document}